\pdfoutput=1
\documentclass[british]{iopart}
\usepackage[T1]{fontenc}
\usepackage[latin9]{inputenc}
\usepackage{units}
\usepackage{amstext}
\usepackage{amssymb}
\usepackage{graphicx}
\usepackage{esint}

\makeatletter

\providecommand{\tabularnewline}{\\}

\usepackage{iopams}
\usepackage{setstack}

\makeatother

\usepackage{babel}
\begin{document}

\title[Correlations in pedestrian flows through
a bottleneck]{Origin of the correlations between exit times in pedestrian flows through
a bottleneck}

\author{Alexandre NICOLAS and Ioannis TOULOUPAS}

\address{LPTMS, CNRS, Univ. Paris-Sud, Universit\'e Paris-Saclay, 91405 Orsay,
France}
\ead{alexandre.nicolas@polytechnique.edu}
\begin{abstract}
Robust statistical features have emerged from the microscopic analysis
of dense pedestrian flows through a bottleneck, notably with respect
to the time gaps between successive passages. We pinpoint the mechanisms
at the origin of these features thanks to simple models that we develop
and analyse quantitatively. We disprove the idea that anticorrelations between
successive time gaps (i.e., an alternation between shorter ones and
longer ones) are a hallmark of a zipper-like intercalation of
pedestrian lines and show that they simply result from the possibility that pedestrians
from distinct `lines' or directions cross the bottleneck within a
short time interval. A second feature concerns the bursts of escapes,
i.e., egresses that come in fast succession. Despite the ubiquity of
exponential distributions of burst sizes, entailed by a Poisson
process, we argue that anomalous (power-law) statistics arise if the
bottleneck is nearly congested, albeit only in a tiny portion of parameter
space. The generality of the proposed mechanisms implies that similar
statistical features should also be observed for other types
of particulate flows.
\end{abstract}

\maketitle

\section{Introduction}

In complex systems, the devil often isn't in the detail, but in the
global response of the system. Indeed, this response may be difficult to predict,
whereas the microscopic dynamics are sometimes more easily grasped. This consideration
notably applies to pedestrian crowds passing through a bottleneck.
In the last fifteen years, this type of flow has been probed intensively
at the scale of individual egresses \cite{isobe2004experiment,hoogendoorn2005pedestrian,kretz2006experimental,nagai2006evacuation,seyfried2009new,seyfried2009empirical,liddle2011microscopic,Pastor2015experimental,garcimartin2016flow,nicolas2017pedestrian,liddle2009experimental}.
Despite the variability of human behaviour and the diversity of morphologies,
robust statistical properties have emerged from the `microscopic'
analysis of the exit time series. Several of the observed features
even turn out to be generic to constricted flows of discrete particles,
but their origins have not been fully elucidated yet.

More precisely, considerable research has focused on the time gaps
$\tau$ between successive exits, whose mean value $\left\langle \tau\right\rangle $
is the inverse of the global flow rate. These time gaps are used to
define diverse quantities, which measure distinct characteristics
of the flow. First, one can study the probability density function
(pdf) $p(\tau)$ of $\tau$. The tail of $p(\tau)$ characterises
the frequency of temporary flow interruptions, usually resulting from
transient clogs. These events get more frequent as the evacuation
is more competitive and the doorway is narrower. In such competitive
cases, the flow is intermittent, consistently with early results of agent-based simulations \cite{helbing2000simulating},
 and $p(\tau)$ is well described by
a power law at large $\tau$ \cite{Zuriguel2014clogging}. The flow intermittency was rationalised 
in a continuous model by Helbing \emph{et al.} \cite{helbing2006analytical} while
the heavy tail of $p(\tau)$ was recently reproduced by one of us within a simple cellular automaton accounting
for the non-uniformity of the pedestrians' behaviours \cite{nicolas2016statistical}. One should nonetheless remark
that similar features have also been observed in e.g. granular flows
through a vibrated hopper \cite{janda2009unjamming}, in which heterogeneities
obviously have a different origin. Interestingly, increasing the pedestrians'
eagerness to escape, i.e., their desired velocities, may stabilise
clogs and thus delay the evacuation \cite{helbing2000simulating,Pastor2015experimental,parisi2007faster}.
Secondly, time gaps $\tau$ are central in the definition of bursts
of escapes, i.e., clusters of pedestrians egressing in rapid succession.
The number $S$ of pedestrians in the burst, called burst size, is
consistently found to follow an exponential distribution \cite{garcimartin2016flow,nicolas2017pedestrian}.
This feature is also observed in constricted flows of grains \cite{mankoc2009role},
sheep \cite{garcimartin2015flow}, and mice \cite{lin2016experimental}.
Finally, one can investigate the correlations between successive time
gaps $\tau$. In diverse experimental settings, marked anticorrelations
have been observed \cite{hoogendoorn2003extracting,seyfried2009new,nicolas2017pedestrian,AlReda2017},
pointing to an alternation between short and long time gaps.

In this work we aspire to gain insight into the origin of the last
two features, namely, the burst sizes and the correlations between
time gaps. For this purpose, comprehensive models of pedestrians dynamics
are of little avail, not only because trustworthy models are still
out of reach, but also because the profusion of details may obscure
the sought elementary causes. Therefore, we adopt an opposite stance
and look for a minimal description that puts in the limelight the
key mechanisms. To do so, after a brief summary of the relevant experimental
results in Section~\ref{sec:Experimental}, we introduce a minimal
model in Section~\ref{sec:Minimal-models-and} and test its ability
to capture the experimental features regarding correlations of time
gaps (Section~\ref{sec:Correlations}) and burst sizes (Section~\ref{sec:Distribution}).

\section{Experimental background and widespread interpretations\label{sec:Experimental}}

Controlled experiments of pedestrian flows through bottlenecks usually
record the series of exit times $t_{1},\ldots,t_{N}$. Using this
time series, the computation of the time gaps $\tau_{p}\equiv t_{p}-t_{p-1}$
for $p=2\ldots N$ is straightforward.

Recalling that, in competitive settings, clogging events cause intermittency,
we can split the outflow into bursts of egresses separated by time
gaps larger than some time scale $\tau^{b}$ (meaning that, within
each burst, $\tau_{p}\leqslant\tau^{b}$). Although the value of $\tau^{b}$
is chosen arbitrarily, the burst sizes $S$ are virtually always found
to be exponentially distributed, regardless of the competitiveness
of the evacuation \cite{garcimartin2016flow,nicolas2017pedestrian}.
This feature extends to granular flows out of a silo in the presence
\cite{mankoc2009role} or in the absence\textbf{ }{\small{}\cite{zuriguel2003jamming}}
of vibrations, sheep entering a barn \cite{garcimartin2015flow},
as well as mice escaping from a smoky chamber \cite{lin2016experimental}.
For completeness, let us mention the only two counterexamples known to
us: Power-law distributed burst sizes were reported in a granular hopper flow through
a rectangular slit \cite{saraf2011power} (presumably because of the existence of diverse arch-formation probabilities,
 depending on the direction of the arch relatively to the slit) and in a study based on a cellular
automaton \cite{perez2002streaming}, although the evidence 
supporting this assertion was not compelling in the latter case.

In granular hopper flows, it is accepted that the exponential tail
in $p(S)$ is a hallmark of the Poisson process governing the formation
of an arch or a vault blocking the exit, which has the same probability
to occur for all exiting grains \cite{mankoc2009role}. In particular, this exponential
decay is encountered in continuous \cite{helbing2006analytical} and discrete
\cite{Masuda2014cellular} models in which the region ahead of the opening is decomposed into semi-circular shells that can
only accommodate a finite number of grains (or pedestrians).  But can the mechanism proposed for grains
be transposed to pedestrian flows? Is the exponential distribution
then truly universal? 

Before we address these questions, let us take a closer look at the
dynamics of egress. To this end, we introduce the correlator 

\[
\mathcal{C}_{j}\equiv\frac{\left\langle \tau_{p+j}\tau_{p}\right\rangle -\left\langle \tau_{p}\right\rangle ^{2}}{\mathrm{Var}(\tau_{p})},
\]
where Var refers to the variance. $\mathcal{C}_{j}$ quantifies the
correlation between a time gap $\tau_{p}$ and the $j$-th next one,
$\tau_{p+j}$. Surprisingly, experiments have shown that $\mathcal{C}_{1}$
is significantly negative, i.e., that short time gaps alternate with
larger ones, at least on average. These anticorrelations were first
observed in cooperative bottleneck flows, in which two lines formed ahead
of the constriction \cite{hoogendoorn2003extracting,seyfried2009new}.
For this reason, they were ascribed to the so called `zipper effect',
sketched in Fig.~\ref{fig:sketch_lanes}a, whereby two, or more,
pedestrian lines get intercalated (like the strands of a
zipper) in the bottleneck, the latter being too narrow to allow pedestrians
to stand shoulder to shoulder. But anticorrelations were then reported
for more competitive evacuations, where the flow was disordered \cite{nicolas2017pedestrian}
(also see the analysis in \cite{AlReda2017}, which used the data
of \cite{garcimartin2016flow}). Since in that case pedestrians did
not line up, the zipper effect could not explain this observation.
Nevertheless, a generalised version of this effect was put forward,
whereby egressing pedestrians compete with, and try to overtake, their
counterparts coming from a \emph{different} direction, which results
in short time gaps, while they maintain a finite headway with pedestrians
walking in the same direction (hence the larger time gaps) \cite{nicolas2017pedestrian}.
At present, these scenarios remain largely pictorial. Therefore, they
need to be validated quantitatively and confronted with alternative
explanations.

\section{Minimal model for bottleneck flow\label{sec:Minimal-models-and}}

\begin{figure}
\noindent \begin{centering}
\includegraphics[width=0.45\columnwidth]{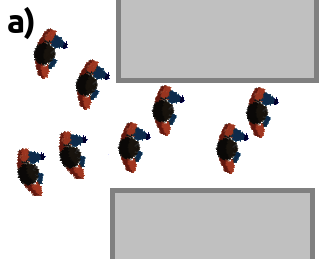}\hspace{1cm}\includegraphics[width=0.45\columnwidth]{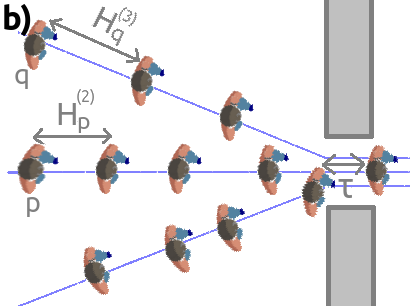}
\par\end{centering}

\caption{\label{fig:sketch_lanes}\textbf{(a)} Sketch of the `zipper effect'
proposed by Hoogendoorn and Daamen \cite{hoogendoorn2005pedestrian}.
\textbf{(b)} Sketch of the situation contemplated in our model, in
the case of $n=3$ lines. The pedestrians' minimal headways $H_{p}^{(k)}$
 are enlarged for clarity.}

\end{figure}

In order to pinpoint the specific elements responsible for the aforementioned
statistical properties, we
develop a minimal model. Pedestrians are assumed to be arranged in
$n$ lanes that converge towards the bottleneck, as sketched in Fig.~\ref{fig:sketch_lanes}b;
their walking speed is set to $v=1$. Within each lane (labelled $k=1\ldots n$),
we denote by $H_{p}^{(k)}(t)$ the minimal time headway in front of
pedestrian $p$. The random variable $H_{p}^{(k)}$ is associated
with the pedestrian's morphology and headway preference; it does not
depend on time. If the bottleneck is permanently congested, there
is no free space and the spacing between pedestrians $p$ and ($p-1$),
measured in time units, is always equal to $H_{p}^{(k)}(t)$; lanes
then consist of contiguous moving blocks of lengths $H_{p}^{(k)}$.

In the most basic version of the model, there is no interaction across
lanes: Pedestrians just walk at constant speed. In a slightly refined
variant, the narrowness of the bottleneck precludes simultaneous passages
through the door: The pedestrian closest to the exit ($p$) exits
first while his or her competitors ($p^{\prime}$) on the \emph{other}
lines ($k^{\prime}$) must halt at a distance $H_{p^{\prime}}^{(k^{\prime})}$
in front of the door until $p$ has crossed the doorway. Note that, with
$n=2$ lines, a zipper effect is obtained in this model variant, if
one imposes that egresses come alternately from the first line and
from the second one.

\section{Correlations between successive time gaps\label{sec:Correlations}}

Using the class of simple models introduced in the previous section,
we now enquire into the `ingredients' that give rise to the empirically
observed alternation between short time gaps and longer ones, i.e.,
$\mathcal{C}_{1}<0$.

\subsection{Single-file model}

First consider a situation with a single line, $n=1$. In this case,
assuming congestion, time gaps $\tau_{p}$ coincide with the headways
$H_{p}$. Can a heterogeneous distribution of headways then give rise
to anticorrelations, $\mathcal{C}_{1}<0$?

At first sight, it is reasonable to assume that the headway $H_{p}$
depends mostly on the intrinsic characteristics of pedestrian $p$.
As pedestrians generally arrived in random order in the experiments,
the headways of successive pedestrians, $H_{p}$ and $H_{p+1}$, are
independent random variables, whence $\left\langle H_{p+1}H_{p}\right\rangle =\left\langle H_{p+1}\right\rangle \left\langle H_{p}\right\rangle =\left\langle H_{p}\right\rangle ^{2}$.
It immediately follows that $\mathcal{C}_{1}=0$; there is no correlation
in this scenario.

Considering the problem in more detail, one realises that, since
$H_{p+1}$ is the spacing between pedestrians $p$ and $p+1$, it
may actually depend on the morphologies of both agents, e.g., their
sizes $S_{p}$ and $S_{p+1}$, because bigger people take more space.
Accordingly, $H_{p+1}$ should grow monotonically with $S_{p}$ and
with $S_{p+1}$. This renders $H_{p}$ and $H_{p+1}$ positively correlated,
because both grow monotonically with $S_{p}$, while $S_{p-1}$ and
$S_{p+1}$ are independent. Hence, one expects $\mathcal{C}_{1}>0$,
contrary to experimental results.

Consequently, the single-file model is unable to account for the observed
anticorrelations. This is consistent with the absence of anticorrelations
in some empirical situations in which pedestrians tended to line up
ahead of the door \cite{nicolas2017pedestrian}. 

Before we turn to multiple-lane models, a few words should be said
about the non-congested case, even though it is less relevant for
the considered experiments \cite{hoogendoorn2005pedestrian,seyfried2009new,garcimartin2016flow,nicolas2017pedestrian}.
In that case, anticorrelations may be obtained if the pedestrians'
walking velocities are not uniform and overtaking is not allowed.
Indeed, the interval ahead of particularly slow pedestrians will grow
with time, whereas the spacing with the pedestrian just behind them
will shrink, and vice versa. Let us mention that anticorrelations
were also reported in a single-lane car traffic model in which the
driver's velocity adjusted not only to the gap in front of it ($p$),
but also to the previous one $(p-1)$, but such anticorrelations were
not observed empirically when single road-lanes were considered \cite{eissfeldt2003effects}.

\subsection{Theoretical results for the two-lane model}

Having discarded the ability of single-file models to capture empirical
observations in congested situations, we turn to a scenario with $n=2$
non-interacting lanes. Let us recall that, ahead of the congested
bottleneck, pedestrians on both lanes ($k\in\left\{ 1,2\right\}$) walk at a constant velocity,
with spacings $H_{p}^{(k)}$ drawn from a random distribution $p_{H}$ that
is assumed independent of the lane $k$. Somewhat counterintuitively, we will
show on the basis of a probabilistic reasoning that this simple model
of \emph{independent} lanes already gives rise to anticorrelated time
gaps ($\mathcal{C}_{1}<0$).

\begin{figure}
\noindent \begin{centering}
\includegraphics[width=0.35\columnwidth]{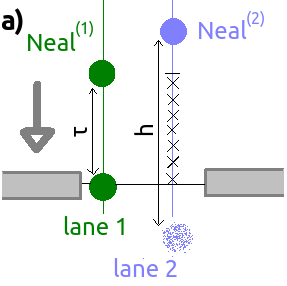}\hspace{1cm}\includegraphics[width=0.35\columnwidth]{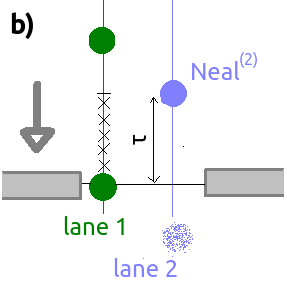}
\par\end{centering}

\caption{\label{fig:sketches_2lanes}Sketches of the situations in which the
next agent to cross the doorway (Neal) is \textbf{(a)} $\mathrm{Neal}^{(1)}$
and \textbf{(b)} $\mathrm{Neal}^{(2)}$. The next time gap is denoted
by $\tau$.}
\end{figure}

Without loss of generality, we can suppose that the $p$-th person 
to egress comes from lane 1. Let $P(\text{Neal}^{(k)},\,\tau)$
be the joint probability that (i) the \textbf{ne}xt \textbf{a}gent
to \textbf{l}eave the room (called Neal) comes from lane $k$ and
(ii) the time gap before this upcoming egress is equal to $\tau$.
To start with, consider $P(\text{Neal}^{(1)},\,\tau)$. The next egress originates
from lane 1 if the next agent on lane 2 (called Neal$^{(2)}$) stands
behind the cross-hatched region in Fig.~\ref{fig:sketches_2lanes}a,
of length $\tau$. Supposing that Neal$^{(2)}$'s headway is $H^{(2)}=h\geqslant\tau$,
the probability to find him out of the hatched region is $\frac{h-\tau}{h}$
(recall that the files are independent). To conclude, we need to estimate
the probability to find such a headway $h$ at the door level. There, a
subtlety arises. Indeed, this probability is not $p_{H}\left(h\right)$,
but $\frac{h}{\left\langle h\right\rangle}p_{H}\left(h\right)$,
because longer headways are encountered more frequently. Bearing in
mind this subtlety, the probability that Neal comes from line 1 reads
\[
\int_{\tau}^{\infty}\frac{h-\tau}{h}\frac{h}{\left\langle h\right\rangle}p_{H}\left(h\right)dh=\int_{\tau}^{\infty}\frac{h-\tau}{\left\langle h\right\rangle}p_{H}(h)dh.
\]
Therefore, 
\begin{equation}
P\left(\text{Neal}^{(1)},\tau\right)=p_{H}\left(\tau\right)\mathcal{I}(\tau),\label{eq:2lanes_P1d}
\end{equation}
where $\mathcal{I}(\tau) \equiv 
\int_{\tau}^{\infty}\frac{h-\tau}{\langle h\rangle}p_{H}\left(h\right)dh
= \int_{\tau}^{\infty} \frac{P(H>h)}{\langle h\rangle}dh$, with $P(H>h)$ the complementary cumulative
distribution function (ccdf) of $p_{H}$.

In the alternative case, sketched in Fig.~\ref{fig:sketches_2lanes}b,
Neal$^{(2)}$ egresses first. This case will occur if Neal$^{(1)}$'s
headway is larger than $\tau$. Thus, it happens with probability
$P(H>\tau)$. We are then left with
the calculation of the probability to find Neal$^{(2)}$ at a distance
$\tau$ ahead of the door. Supposing that Neal$^{(2)}$'s headway
is $h\geqslant\tau$, this probability is simply $1/h$, because the
doorway can intersect Neal$^{(2)}$'s headway anywhere, owing to the
independence between lines. Recalling the aforementioned subtlety,
we arrive at 
\begin{eqnarray}
P\left(\text{Neal}^{(2)},\tau\right) & = & P\left(H>\tau\right)\int_{\tau}^{\infty}\frac{1}{h}\frac{h}{\left\langle h\right\rangle}p_{H}\left(h\right)dh\nonumber \\
 & = & P\left(H>\tau\right)\frac{P\left(H>\tau\right)}{\left\langle h\right\rangle}.\label{eq:2lanes_P2d}
\end{eqnarray}

Equations~\ref{eq:2lanes_P1d}-\ref{eq:2lanes_P2d} express the probabilities
to find a time gap $\tau_{p}$, if we only know that the last person
who egressed came from lane 1. In order to evaluate the correlator
$\mathcal{C}_{1},$ we also need to estimate $\tau_{p+1}$, i.e.,
to consider not only the agent who egresses next (Neal), but also
the \textbf{se}cond next \textbf{a}gent to \textbf{l}eave, whom we
name Seal. The notations deserve a remark: $\tau_{p+1}$ is the time
gap measured just after $\tau_{p}$, regardless of the line from which
Seal comes. That being said, the derivation of the joint probabilities
$P\left(\text{Neal}{}^{(k)},\,\tau_{p},\,\text{Seal}{}^{(k^{\prime})},\,\tau_{p+1}\right)$,
i.e., Neal comes from line $k$ and egresses after $\tau_{p}$, followed
after $\tau_{p+1}$ by someone from line $k^{\prime}$, is very similar
to the derivation detailed above for $\tau_{p}$ and leads to the
following expressions,

\begin{eqnarray}
P\left(\text{Neal}^{(1)},\,\tau_{p},\,\text{Seal}^{(1)},\,\tau_{p+1}\right)=p_{H}\left(\tau_{p}\right)\,p_{H}\left(\tau_{p+1}\right)\mathcal{I}\left(\tau_{p}+\tau_{p+1}\right) \label{eq:2lanes_P_12}
\\
P\left(\text{Neal}^{(1)},\,\tau_{p},\,\text{Seal}^{(2)},\,\tau_{p+1}\right)=p_{H}\left(\tau_{p}\right)\,P\left(H>\tau_{p+1}\right)\frac{P\left(H>\tau_{p}+\tau_{p+1}\right)}{\left\langle h\right\rangle} \nonumber
\\
P\left(\text{Neal}^{(2)},\,\tau_{p},\,\text{Seal}^{(1)},\,\tau_{p+1}\right)=p_{H}\left(\tau_{p}+\tau_{p+1}\right)\frac{P\left(H>\tau_{p}\right)}{\left\langle h\right\rangle}P\left(H>\tau_{p+1}\right) \nonumber
\\
P\left(\text{Neal}^{(2)},\,\tau_{p},\,\text{Seal}^{(2)},\,\tau_{p+1}\right)=P\left(H>\tau_{p}+\tau_{p+1}\right)\frac{P\left(H>\tau_{p}\right)}{\left\langle h\right\rangle}p_{H}\left(\tau_{p+1}\right). \nonumber
\end{eqnarray}

Finally, these two-point probabilities are inserted into the time
gap correlator, viz., $\mathcal{C}_{j}=\left\langle \delta\tau_{p}\delta\tau_{p+1}\right\rangle /\left\langle \delta\tau_{p}^{2}\right\rangle $,
where 

\begin{equation}
 \hspace*{-2.3cm}
\left\langle \delta\tau_{p}\delta\tau_{p+1}\right\rangle   = \iint_{0}^{\infty} (\tau_{p}-\bar{\tau})(\tau_{p+1}-\bar{\tau})
\sum_{k,k^{\prime}=1,2}
P\left(\text{Neal}^{(k)},\tau_{p},\text{Seal}^{(k^{\prime})},\tau_{p+1}\right)d\tau_{p}d\tau_{p+1}.\label{eq:2lane_correlations_gal}
\end{equation}
Here, the average time gap $\bar{\tau}$ obeys $\bar{\tau}=\langle h\rangle / 2$
and the fluctuation $\delta\tau_{p}$ is defined as $\tau_{p}-\bar{\tau}$.

\subsection{Evaluation of the correlator for specific headway distributions}

Unfortunately, Eq.~\ref{eq:2lane_correlations_gal} is complex and
the sign of $\mathcal{C}_{1}$ does not immediately transpire from
its inspection. Therefore, we will now focus on specific headway distributions,
which will simplify the formulae and allow us to test them against
simulations of the model.

To start with a simple, but enlightening, example, we assume that
headways are constant, i.e., $p_{H}\left(h\right) = \delta\left(h-\bar{h}\right)$. This comes down
to studying a perfect zipper in which two regularly ordered lines intercalate with a random offset.
In this case, $\bar{\tau}=\bar{h} / 2$ and Eqs~\ref{eq:2lanes_P_12}-\ref{eq:2lane_correlations_gal}
yield 
\begin{equation*}
\left\langle \delta\tau_{p}^{2}\right\rangle  =  \frac{\bar{h}^{2}}{12}\text{ and }
\left\langle \delta\tau_{p}\delta\tau_{p+1}\right\rangle   =  \frac{-\bar{h}^{2}}{12}.
\end{equation*}
It follows that $\mathcal{C}_{1}=-1$: Successive time gaps are thus
anticorrelated, despite the independence of the lines. There is an
intuitive explanation to this observation. Consider an interval between
pedestrians in line 1. Because of the random offset between the lines,
a pedestrian in line 2 generally splits this interval into a small
section and a large one. Since spacings are equal in the two lines,
these small and large intervals alternate cyclically, thus generating
anticorrelations.

This reasoning hinges on the regular ordering of pedestrians on both
lanes. What will remain of it, should the flow be more disordered and pedestrians
unequally spaced? To test this more realistic scenario, we consider
Gaussian distributions of headways $p_{H}(h)=\left(\delta h\sqrt{2\pi}\right)^{-1}\exp\left[\frac{\left(h-\bar{h}\right)^{2}}{2\delta h^{2}}\right]$,
where $\bar{h}=1$ is the mean, and the standard deviation $\delta h$
will be varied to study the influence of the level of disorder. In
practice, it will remain small enough for occurrences of $h<0$ to be
negligible (in the numerical simulations, negative values of $h$
are shifted to zero). With such Gaussian distributions, explicit analytical
formulae can be obtained for the two-point probabilities of Eqs.~\ref{eq:2lanes_P_12},
by inserting the following equalities, 
\[
P\left(H>h\right)=\frac{1}{2}\mathrm{erfc}\left(\frac{h-\bar{h}}{\delta h\sqrt{2}}\right),\text{ and}
\]
 
\[
\mathcal{I}(\tau)=\frac{\delta h^{2}}{\bar{h}}p_{H}(\tau)+\frac{\bar{h}-\tau}{2\bar{h}}\mathrm{erfc}\left(\frac{\tau-\bar{h}}{\delta h\sqrt{2}}\right),
\]
where $\mathrm{erfc}$ is the complementary error function. We then
rely on numerical integration to evaluate the correlator of Eq.~\ref{eq:2lane_correlations_gal}.
The correlations $\mathcal{C}_{1}$ resulting from this integration
accurately match the results of direct numerical simulations of the
model for small $\delta h$, as evidenced in Fig~\ref{fig:C1_vs_sigma}(left).
For larger $\delta h$, the truncation of of $p_{H}$ to positive
$h$ in the numerics causes minor differences. Most importantly, we
see that $\mathcal{C}_{1}$ is significantly negative not only for
vanishing $\delta h$ (almost equal spacings), but also when the flow
is entirely disordered, with a rather broad distribution of headways
($\delta h/\bar{h}>\nicefrac{1}{2}$). We have checked that this also holds
if the headway distributions differ between the two pedestrian lines (i.e., if their
parameters $\bar{h}$ or $\delta h$ differ).
To rephrase this noteworthy result, there tends to
be an alternation between short time gaps and longer ones even if
pedestrians are not regularly spaced and the lines do not interact. 

As a corollary, anticorrelations ($\mathcal{C}_{1}<0$) bring no evidence
of a zipper-like intercalation of lines imposed by the narrowness
of the bottleneck .

\subsection{Effect of extensions of the model\label{sub:Model-extensions-and}}

\begin{figure}
\noindent \begin{centering}
\includegraphics[width=0.49\columnwidth]{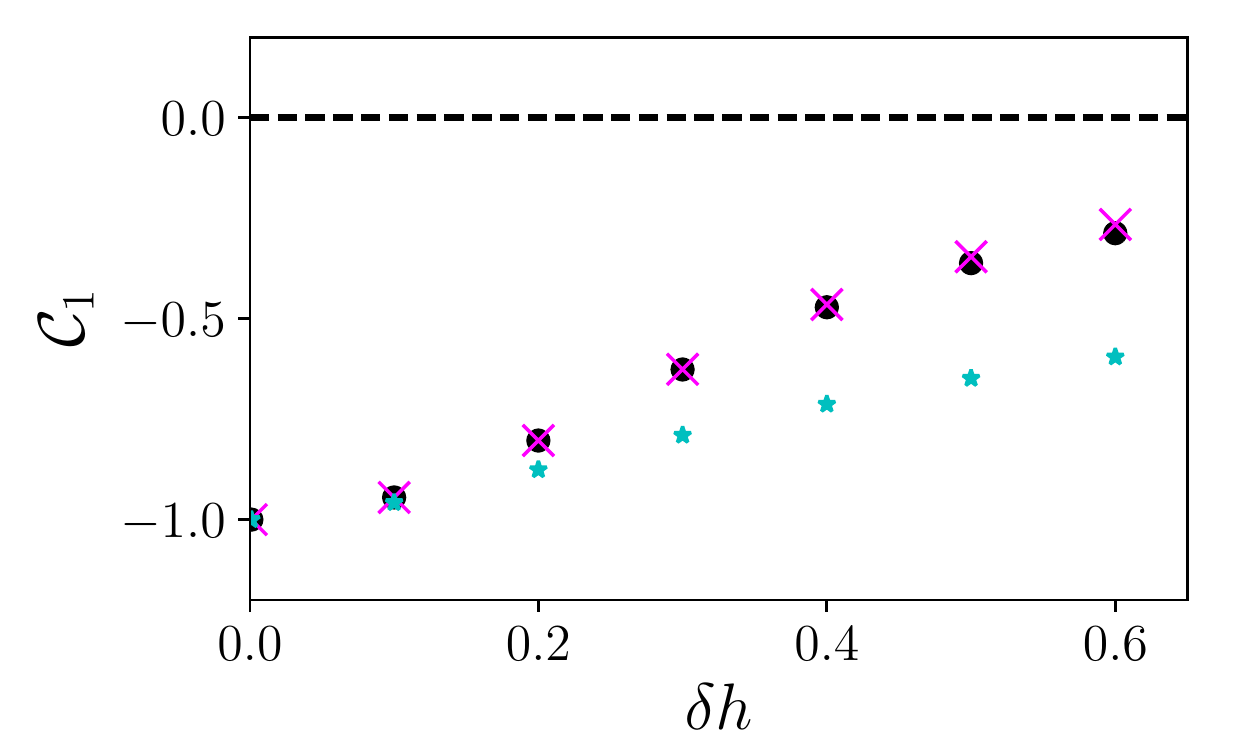}~\includegraphics[width=0.49\columnwidth]{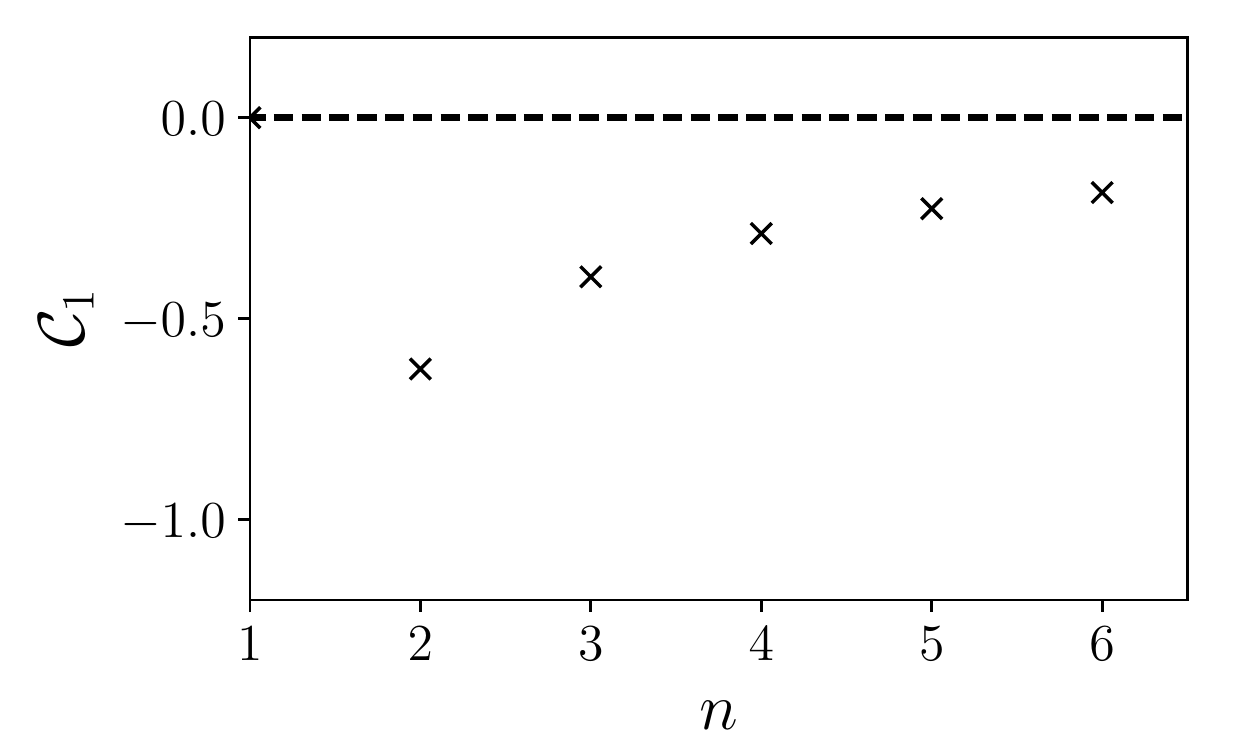}
\par\end{centering}

\caption{\label{fig:C1_vs_sigma}\emph{Correlations between successive time
gaps}, $\mathcal{C}_{1}=\frac{\left\langle \delta\tau_{j}\delta\tau_{j+1}\right\rangle }{\left\langle \delta\tau_{j}^{2}\right\rangle }$, 
for Gaussian distributions of minimal headways with unit mean and
standard deviation $\delta h$.
(\emph{Left}) Dependence of $\mathcal{C}_{1}$ on $\delta h$ for
two independent lanes of pedestrians. Black dots were obtained
with the theoretical expression of Eq.~\ref{eq:2lane_correlations_gal};
pink crosses are simulation results. Cyan stars correspond to simulations
of the model enforcing an alternation between lines.
(\emph{Right}) Variations of $\mathcal{C}_{1}$ with the number $n$
of (independent) lanes, for $\delta h=0.3$.}
\end{figure}

Admittedly, the simple model presented above does not account for
all observed features. For instance, the experiments of \cite{nicolas2017pedestrian}
unveiled a general alternation in the incident directions of egressing
pedestrians, i.e., between people coming from the left and from the
right. This feature can be enforced in the model by making $\mathrm{Neal}^{(1)}$
wait for $\mathrm{Neal}^{(2)}$ to exit if the last egress was from
lane 1; then, $\mathrm{Neal}^{(1)}$ egresses immediately after $\mathrm{Neal}^{(2)}$.
This systematic alternation between lines brings our model closer
to that of Hoogendoorn and Daamen \cite{hoogendoorn2005pedestrian},
in which headways were constrained by the impossibility of cross-line
overtaking inside the bottleneck. Numerical simulations, represented by
cyan stars on Fig.~\ref{fig:C1_vs_sigma}(left), show that this new
constraint further enhances the anticorrelations, that is, make $\mathcal{C}_{1}$
more negative. This is easily understood: All scenarios $\left(\text{Neal}{}^{(1)},\,\tau_{p},\,\text{Seal}{}^{(1)},\,\tau_{p+1}\right)$
in the previous model turn into scenarios with a longer $\tau_{p}$
and $\tau_{p+1}=0$.

Stronger interactions between lines are expected if the door is so
narrow that pedestrians can only pass one by one, which means that
the two lines effectively merge at the entrance of the bottleneck.
This is incorporated into our model by making pedestrian $p$ (on
lane $k$) wait at a distance $H_{p}^{(k)}$ ahead of the door whenever
somebody from the other line is currently passing through the door.
The next agent to egress is the one that stands closer to the door
and the duration of his or her passage is equal to this distance to the door
(recall that $v=1$). It turns out that this one-by-one passage rule
totally suppresses the time-gap anticorrelations (\emph{numerical
data not shown}); this is due to the fact that agents on
lane 2 no longer split the headways on lane 1. As one can imagine,
prescribing a systematic alternation between lines, as in a \emph{bona
fide} zipper, does not restore the anticorrelations. 

On the contrary, larger doors can accommodate multiple virtually independent
lanes. Restoring the independence between lanes, we compute the variations
of $\mathcal{C}_{1}$ with the number of lanes $n$. The results are
plotted in Fig.~\ref{fig:C1_vs_sigma}(right) and prove that anticorrelations
persist for $n\geqslant2$, even though they weaken with increasing
$n$.

To summarise the results of this section, within the rather general
framework of our model, time-gap anticorrelations emerge if several
pedestrians (from distinct lines or directions) can cross the door
within a very short interval, as compared to the headways. Accordingly,
we speculate that anticorrelations would be suppressed in any experiment
with a very narrow door, only allowing one pedestrian to pass at a
time.

\section{Distribution of burst sizes\label{sec:Distribution}}

When the time-series of exits is inspected at a more global scale
than that of individual time gaps, clusters of uninterrupted escapes,
called bursts, become apparent in competitive conditions \cite{Zuriguel2014clogging,Pastor2015experimental,nicolas2017pedestrian}.
In practice, a more precise criterion is required to delimit bursts.
Usually, one sets an arbitrary upper bound $\tau^{b}$ on the admissible
time-gaps within a burst. A similar criterion can be applied to identify
bursts in more ordered flows.

In Section~\ref{sec:Experimental}, we recalled that, regardless
of the precise value $\tau^{b}$, the competitiveness of the crowd
and the experimental settings, pedestrian experiments consistently
measured exponentially distributed burst sizes $S$, at large $S$
\cite{garcimartin2016flow,nicolas2017pedestrian}. The model
introduced in Section~\ref{sec:Minimal-models-and} and all variants
discussed so far are no exception to this rule, as confirmed by the
data shown in Fig~\ref{fig:exp_burst_sizes}. Is the exponential
pdf of burst sizes then a truly universal feature of bottleneck flows?
In the following, after examining the origin of this feature, we contemplate
situations which lead to heavier-than-exponential distributions of
$S$.

\subsection{A Poisson process in the congested case\label{sub:A-Poissonian-process}}

\begin{figure}
\noindent \begin{centering}
\includegraphics[width=0.5\columnwidth]{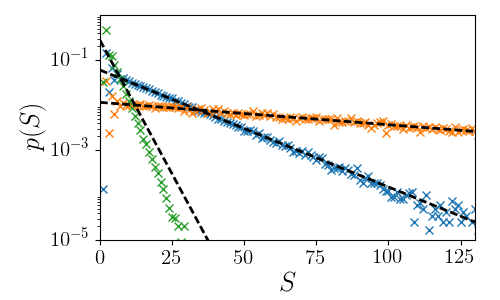}
\par\end{centering}

\caption{\label{fig:exp_burst_sizes}Distributions of burst sizes $S$ obtained
in simulations of $n=2$ non-interacting lanes with Gaussian-distributed
headways (mean $\bar{h}=1$, standard deviation $\delta h=0.3$),
with the burst criterion set to $\tau^{b}=0.7$ (\emph{green crosses}),
1 (\emph{blue crosses}), 1.2 (\emph{orange crosses}). The dashed lines
are the exponential fits $p(S)=p_{c}e^{-p_{c}S}$, where $p_{c}$
is evaluated using Eqs.~\ref{eq:2lanes_P1d}-\ref{eq:2lanes_P2d}.
Note the semi-logarithmic axes.}

\end{figure}

As long as the probability that a burst is interrupted before the passage
of a new agent takes a \emph{constant} value $p_{c}$, with $0<p_{c}<1$,
the egress will be a Poisson process of rate $p_{c}$, which implies
$p(S)\approx p_{c}e^{-p_{c}S}$ for $S\rightarrow\infty$. In a granular
hopper flow, $p_{c}$ is the clogging probability per grain \cite{mankoc2009role}.
In our pedestrian flow models, $p_{c}$ simply represents the probability
to find a time gap larger than $\tau^{b}$, $p_{c}=P\left(\tau>\tau^{b}\right)$.
Despite the presence of short-term correlations $\mathcal{C}_{p}$
(which undermine the argument), evaluating $p_{c}$ in this way with
the help of Eqs.~\ref{eq:2lanes_P1d}-\ref{eq:2lanes_P2d} results
in very good fits to simulation results, as shown in Fig.~\ref{fig:exp_burst_sizes},
provided that $\tau_{b}$ is sufficiently large. For smaller $\tau_{b}$,
$p(S)$ seems to still decay exponentially at large $S$, but the
correlations between successive time gaps induce deviations
from the fit based on $p_{c}$.

\subsection{Possibility of anomalous statistics}

Despite all apparences to the contrary, we argue that exponential burst size distributions
are not universal and that power-law distributions
can arise in particular situations close to congestion. 
Extending the model presented in Section~\ref{sec:Minimal-models-and}
to non-congested situations, we denote by $X_{p}^{(k)}(t)\leqslant0$
the position of pedestrian $p$ on lane $k$ at time $t$, with the
origin at the door. The agent's \emph{actual} headway 
\[
\delta X_{p}^{(k)}(t)\equiv X_{p-1}^{(k)}(t)-X_{p}^{(k)}(t)
\]
is always larger than the \emph{minimal} headway $H_{p}^{(k)}$, viz.,
$\delta X_{p}^{(k)}(t)\geqslant H_{p}^{(k)}$. Both $H_{p}^{(k)}$
and the initial headway $\delta X_{p}^{(k)}(0)$ are random variables
drawn once and for all from distributions that are identical for all
lanes (i.e., that do not depend on $k\in\left\{ 1,\ldots,n\right\} $).
To illustrate our findings, we will choose Gaussian distributions,
with respective means $\left\langle H_{p}^{(k)}\right\rangle =\overline{H}$
and $\left\langle \delta X_{p}^{(k)}(0)\right\rangle =n\,\overline{\delta X}$,
where $\overline{\delta X}$ is the average initial spacing between
pedestrians, all lanes combined.

The door is supposed narrow, so that only one agent (Neal) can pass
at a time, as in the variant introduced in Section~\ref{sub:Model-extensions-and}.
We remind the reader that during Neal's passage, his competitors $p$
on lane $k$ cannot move closer than a distance $H_{p}^{(k)}$ to
the door, which may force them to halt and may reduce the headways
$\delta X_{p^{\prime}}^{(k)}(t)$ of the people behind. As soon as
Neal has left, the agent closest to the door can egress.

Quite interestingly, numerical simulations show that for a (limited)
range of parameters this model yields distributions of burst sizes
$p(S)$ whose tails are heavier than exponential and are well fitted
by power laws (see Fig.~\ref{fig:power_law_ccdf}).

\subsection{Delay accumulation and power-law statistics}

To clarify the origin of these anomalous statistics, a theoretical
analysis of the model is helpful. The analysis is simplified by focusing
on the line that results from the merger of the $n$ lines at the
door. Therefore, we label pedestrians according to their order of
egress and drop the $(k)$ superscripts referring to the lane from
which they come. With this new labelling, the $\delta X_{p}(t)$'s
are algebraic distances and may be positive or negative. Besides,
the sequence of time gaps between egresses is $\tau_{1},\ldots,\tau_{N}$
and pedestrians $(p-1)$ and $p$ belong to the same burst if $\tau_{p}\leqslant\tau^{b}$. 

If the bottleneck is not congested, the time gaps $\tau_p$ need not coincide
with the minimal headways $H_{p}$, but result from the shrinkage
of the interpedestrian spacings due to halts. We are thus led to study
the propagation of halts (or <<jams>>). Let $D_{p}$ be the cumulative
halting time (delay) of agent $p$ from start $(t=0)$ to egress
$(t=t_{p})$. To find an iterative relation between $D_{p}$ and $D_{p-1}$,
we consider a fictional situation in which pedestrians just walk at
constant velocity ($v=1$), without halts; this can be realised by
setting $H_{p}=0$ for all agents $p$, all other things being equal.
In the fictional scenario, referred to by prime superscripts, spacings
are preserved, viz., 
\[
\delta X_{p}^{\prime}(t)=\delta X_{p}^{\prime}(0)=\delta X_{p}(0),
\]
Recalling that $X_{p}(t_{p})=0$, we remark that the delay $D_{p}$
in the actual model coincides with $X_{p}^{\prime}(t_{p})$, so that

\begin{eqnarray*}
D_{p} & = & X_{p}^{\prime}(t_{p})\\
 & = & X_{p}^{\prime}(t_{p-1})+t_{p}-t_{p-1}\\
 & = & X_{p-1}^{\prime}(t_{p-1})-\delta X_{p}(0)+\tau_{p}\\
 & = & D_{p-1}-\delta X_{p}(0)+\tau_{p}.
\end{eqnarray*}
Now, $D_{p}>0$ means that agent $p$ has halted due to congestion,
which reduced his or her actual headway to $H_{p}$, so $\tau_{p}=H_{p}$.
The alternative case, $D_{p}=0$, corresponds to only free walk for
agent $p$ and implies $\tau_{p}=-D_{p-1}+\delta X_{p}(0)$. In Table~\ref{tab:Coupled_evolutions},
we sum up the coupled equations of evolution of $\tau_{p}$ and $D_{p}$
that directly follow from these considerations. 

\begin{table}
\noindent \begin{centering}
\begin{tabular}{c|c|c}

 & {\bf If $\boldsymbol{D_{p-1}>H_{p}-\delta X_{p}(0)}$} (congestion),  & {\bf Otherwise} (free walk),\tabularnewline
\hline
$\tau_{p}=$ & $H_{p}$ & $-D_{p-1}+\delta X_{p}(0)$\tabularnewline

$D_{p}=$ & $D_{p-1}+H_{p}-\delta X_{p}(0)$ & 0\tabularnewline

\end{tabular}
\par\end{centering}

\caption{\label{tab:Coupled_evolutions}Iterative relations governing the coupled
evolutions of time gaps $\tau_{p}$ and cumulative delays $D_{p}$.}
\end{table}

Since $H_{p}$ and $\delta X_{p}(0)$ are random variables with mean
values $\overline{H}$ and $\overline{\delta X}$, respectively, $D_{p}$
performs a one-dimensional random walk in ``time'' $p$ with a bias
$\overline{H}-\overline{\delta X}$, close to the absorbing boundary
$0^{+}$. Depending on the bias, different flow regimes are expected.
(i) If the bias is strongly positive, i.e., $\overline{\delta X}\ll\overline{H}$
, congestion is expected for the greatest part of time. (ii) For $\overline{H}\simeq\overline{\delta X}$,
$D_{p}$ performs a virtually unbiased random walk. The survival times
$S_{\mathrm{cong}}$ of the congested phases $D_{p}>0$ are thus distributed
according to $p(S_{\mathrm{cong}})\sim S_{\mathrm{cong}}^{-\frac{3}{2}}$.
(iii) For $\overline{\delta X}\gg\overline{H}$, free walk prevails,
and the distribution of congested durations $S_{\mathrm{cong}}$ will
decay quickly at large time scales, due to the downward drift of $D_{p}$,
of magnitude $\overline{H}-\overline{\delta X}$. 

\begin{figure}
\noindent \begin{centering}
\includegraphics[width=0.49\columnwidth]{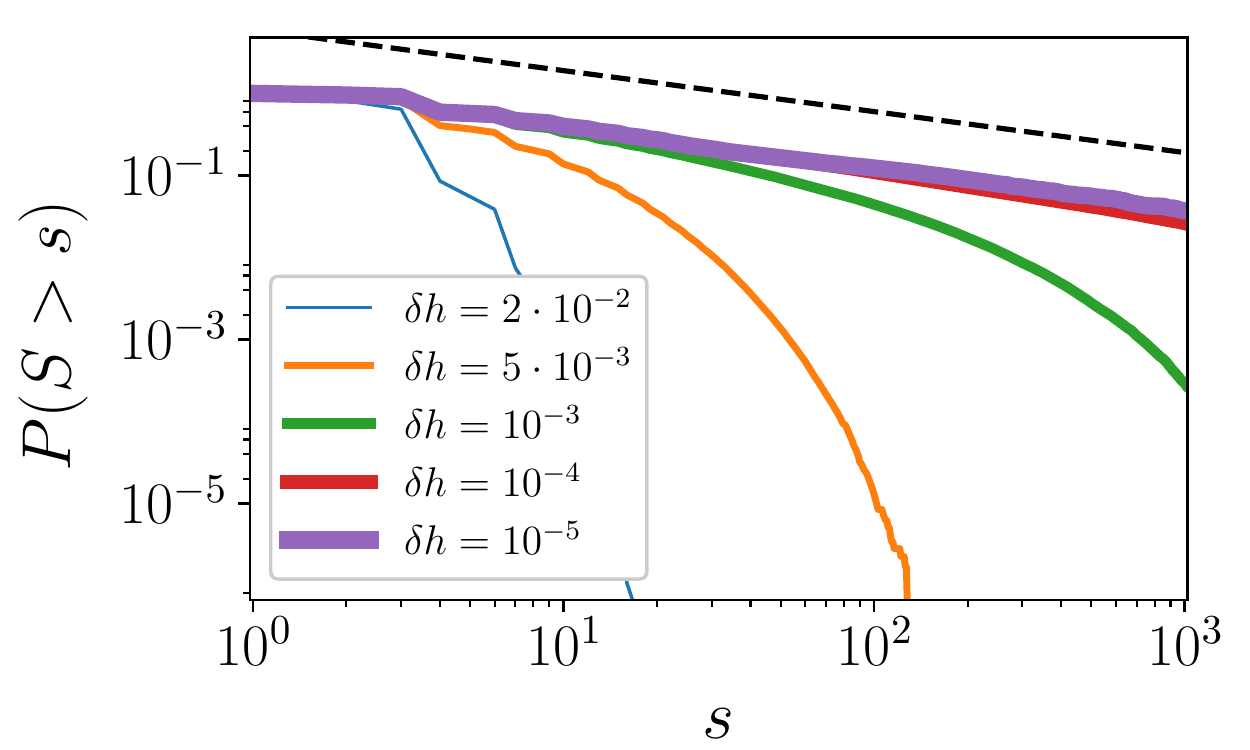}~\includegraphics[width=0.49\columnwidth]{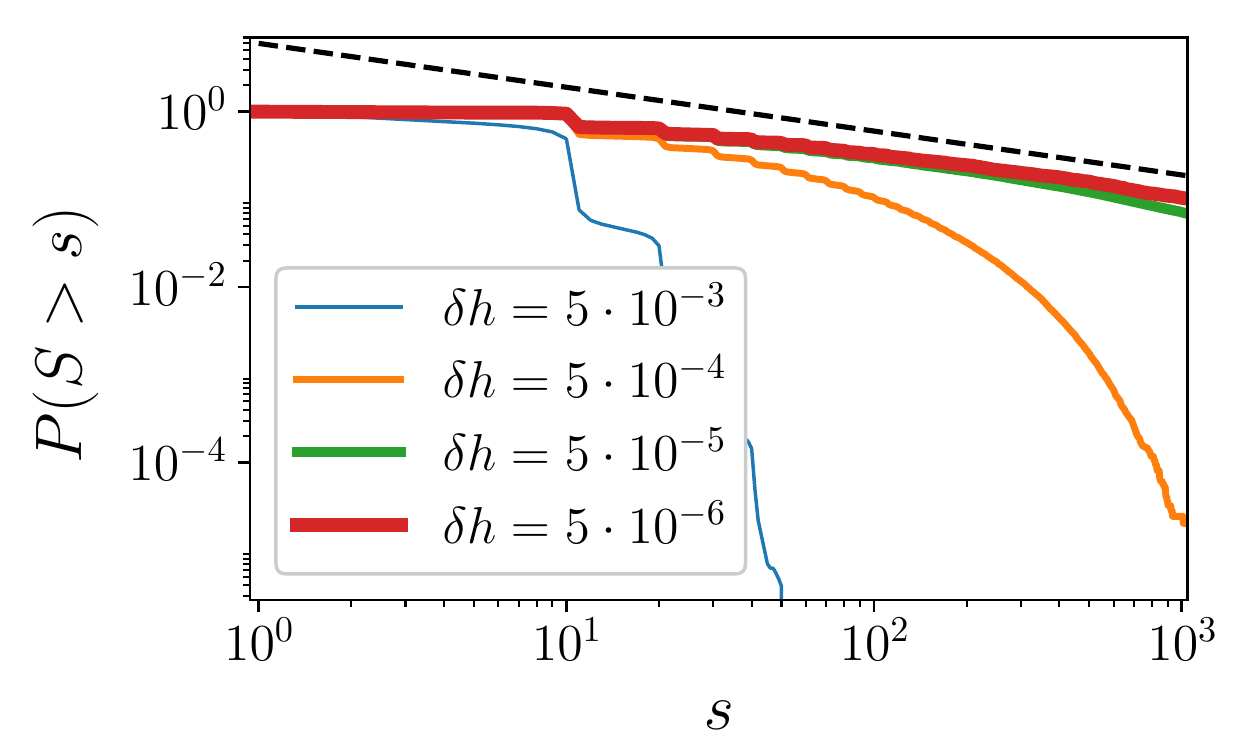}
\par\end{centering}

\caption{\label{fig:power_law_ccdf}Complementary cumulative distributions $p(S)$
of burst sizes $S$ obtained numerically for Gaussian distributions
of minimal headways of mean $\overline{H}=1$ and standard deviation
$\delta h$, as indicated in the legend. $n=3$ lanes (left)
and $n=10$ lanes (right). The initial distribution of spacings is
also Gaussian, with mean value $n(1+10\delta h)$. The burst criterion $\tau^b$ was adjusted to be the average between 
$\overline{H}$ and the mean initial spacing (with all lines combined).
 The dashed black lines have slopes
$\nicefrac{-1}{2}$, which corresponds to an exponent $\nicefrac{-3}{2}$
for the pdf.}

\end{figure}

In Section~\ref{sub:A-Poissonian-process}, we saw that bursts within
a congested phase do not display anomalous (heavy-tailed) statistics.
Therefore, to have a chance to observe anomalous statistics, bursts
must not end during these phases. Thus, $P\left(H>\tau^{b}\right)$
must be negligible. With a somewhat sloppy notation, we write this
condition $\sup\left(H\right)<\tau^{b}$. At the same time, $\tau^{b}$
must be small enough to separate bursts between subsequent congested
phases, viz., $\tau^{b}\leqslant\overline{\delta X}$ (otherwise,
bursts would follow a Poisson process dependent on the initial spacings).
Under these combined conditions, the burst sizes coincide with the
survival times $S_{\mathrm{cong}}$, which are power-law distributed
in regime (ii), $\overline{H}\simeq\overline{\delta X}$. Figure~\ref{fig:power_law_ccdf}
shows that, if all these conditions are fulfilled, a power-law distribution
of burst sizes with exponent $-\nicefrac{3}{2}$ is indeed observed.

However, this regime is highly constrained in parameter space, since
the inequality $\sup\left(H\right)<\tau^{b}\leqslant\overline{\delta X}$
must be fulfilled at the same time as $\overline{H}\simeq\overline{\delta X}$.
In practice, this is only achieved if the minimal headways $H_{p}$
are quite narrowly distributed and $\tau^{b}\simeq\sup\left(H_{p}\right)\simeq\overline{\delta X}$
. Otherwise, $p(S)$ displays a cut-off.

On a more subtle note, it may be remarked that the above reasoning
is oblivious to possible correlations in the random walk of $D_{p}$
due to the lane structure of the flow. It discards the fact that,
for agents on the same lane, $H_{p}^{(k)}-\delta X_{p}^{(k)}(0)$
is always negative. This fact enhances the likelihood of the interruption
of a congested phase every $n$ egresses, which leads to the pdf (and
ccdf) oscillations of period $n$ that can be seen in Fig.~\ref{fig:power_law_ccdf}.

\subsection{Analogy with fibre bundles and depinning problems}

There is perhaps a more intuitive way to interpret the distributions
of burst sizes $S$. It consists in drawing a parallel with fibre
bundle models \cite{sornette1989self} or depinning models \cite{fisher1983threshold}.
In these models, the rupture of a fibre, or the  motion of a site, destabilises 
 other elements via the redistribution of some positive load $\tau$, which may
trigger another rupture. In the mean-field picture, the redistributed load is 
equally shared among all other elements. To clarify the analogy with our models,
we liken the egress of a pedestrian to the rupture
of a fibre. In the same way as a broken fibre enhances the probability of subsequent rupture,
an egress postpones the end of a burst. The extra time thus given to other agents
to exit in the
same burst is equal to $\tau=\tau_{p}+\min(\tau_{p+1},\tau^{b})>0$.

Now, it is known that fibre bundle models display cascades
of ruptures whose sizes $S$ are power-law distributed close to criticality, with $p(S)\sim S^{-\frac{3}{2}}$
\cite{sornette1989self}. Farther from the critical point, \emph{e.g.}, when the system is too stable
to be entirely destabilised by a single rupture, the power-law
distribution is cut-off exponentially. In the light of this analogy, we should not have been surprised to detect a  power-law distribution of 
burst sizes with exponent $3/2$ in the pedestrian model at the transition between a
dilute flow and a congested one, provided that the criticality of this transition was not washed out by the disorder in
the headways and the finite number of lanes.

\section{Conclusions}

In summary, we have enquired into the microscopic dynamics of pedestrian
flows through a bottleneck, with a strong emphasis on the time gaps
$\tau$ between egresses. Our study, based on simple models, sheds
light on the mechanisms underlying the experimentally evidenced statistical
properties of these time gaps. More precisely, our models show that,
contrary to a widespread belief, the anticorrelations between successive
time gaps (that is, the alternation between short and long time gaps),
which have been reported in various settings, are not a hallmark of
a zipper-like intercalation of pedestrian lines. Instead, they naturally
emerge whenever the bottleneck is wide enough to allow two or more
pedestrians to cross it within a short time interval. Under these
conditions, the pedestrian's headways within a line are split into
(generally unequal) sections by the independent passage of another
pedestrian, thereby yielding anticorrelations. The minimality of these
conditions rationalises the observation of this effect in competitive
evacuations as well, where pedestrians did not line up in front of
the door. Turning to the second statistical feature, the ubiquitous
exponential distributions of burst sizes owe their origin to the constant
probability that the lagging egress of a pedestrian marks the end
of the burst. Nevertheless, our study unveiled the possibility of
anomalous (power-law) statistics if the system is on the brink of
congestion and the burst criterion is fine-tuned; the model can
then be amalgamated to a fibre bundle. However, this regime
only covers a tiny portion of parameter space. It remains to be seen
if some vestige of these anomalous statistics can be detected experimentally,
in the form of a power law with an exponential cut-off.

Finally, it should be stressed that the mechanisms giving rise to
these statistical properties are very general. In fact, they do not
rely on any distinctive feature of pedestrian motion. Therefore, we
also expect anticorrelations between time gaps to be observed in other
types of constricted flows too, such as hopper flows of granular discs,
evacuations of mice or multiple-lane vehicular traffic.

\ack
IT's internship was funded by a grant of the French Labex PALM (ANR-11-IDEX-0003-02, Perce-Foule). We thank C\'ecile Appert-Rolland for multiple discussions.

\providecommand{\newblock}{}

\end{document}